\documentclass[12pt]{article}
\usepackage{amsfonts}
\usepackage{epsfig,amssymb,euscript}
\usepackage{amsmath,amscd}
\usepackage[usenames,dvipsnames]{color}
\usepackage[pdftex,unicode,implicit]{hyperref}
\hypersetup{%
  pdftitle    = {Einstein-Weyl structures and de Sitter Supergravity},
  pdfkeywords = {classification of d=5 fSUGRA} {d=5 fakeSupergravity with Abelian vector multiplets} {Einstein-Weyl} {Gauduchon-Tod},
  pdfsubject  = {hep-th},
  pdfauthor   = {J.~B.~Gutowski, A.~Palomo-Lozano, W.~A.~Sabra},
  pdfcreator  = {pdf\LaTeXe\ with package \flqq hyperref\frqq},
  pdfproducer = {pdf\LaTeXe\ with package \flqq hyperref\frqq},
  pdfpagemode = None,  %change this to FullScreen if you want...
  pdffitwindow= true,  % Open the pdf file without the sidebars open.
  unicode     = true,
  plainpages  = false,
  colorlinks  = false,  % links have no different colors..
 }

\numberwithin{equation}{section}
\newcommand{\be}{\begin{eqnarray}}
\newcommand{\ee}{\end{eqnarray}}
\newcommand{\bea}{\begin{eqnarray}}
\newcommand{\eea}{\end{eqnarray}}
\newcommand{\ba}{\begin{array}}
\newcommand{\ea}{\end{array}}
\newcommand{\nn}{\nonumber \\}

\newcommand{\bR}{\mathbb{R}}

\def\cL{{\cal L}}

\def\bbe{{\bf{e}}}

\begin{document}
\begin{titlepage}
\vfill
\begin{flushright}
IFT-UAM/CSIC-11-40
\end{flushright}
\vfill
\begin{center}
   \baselineskip=16pt
  {\Large\bf Einstein Weyl Structures and de Sitter Supergravity}
   \vskip 2cm
       Jan B. Gutowski$^1$, Alberto Palomo-Lozano$^2$
    and W. A. Sabra$^3$\\
 \vskip .6cm
      \begin{small}
      $^1$\textit{Department of Mathematics, King's College London.\\
      Strand, London WC2R 2LS\\United Kingdom \\
        E-mail: jan.gutowski@kcl.ac.uk}
        \end{small}\\*[.6cm]
          \begin{small}
      $^2$\textit{ Instituto de F\'isica Te\'orica UAM/CSIC\\
	C/ Nicol\'as Cabrera 13-15\\
      Ciudad Universitaria de Cantoblanco, E-28049 Madrid, Spain   \\
        E-mail: alberto.palomo@uam.es}
\end{small}\\*[.6cm]
    \begin{small}
      $^3$\textit{Centre for Advanced Mathematical Sciences and
        Physics Department, \\
        American University of Beirut, Lebanon \\
        E-mail: ws00@aub.edu.lb}
        \end{small}
   \end{center}
\vfill
\begin{center}
\textbf{Abstract}\\
\vspace{2ex}
\begin{minipage}{14cm}
The geometric structure of the null solutions of  de Sitter D=5 gauged supergravity coupled 
to vector multiplets is investigated. These solutions are  Kundt metrics, 
constructed from a one-parameter family of three dimensional Gauduchon-Tod base spaces. 
We give examples, including the near horizon geometries previously found in \cite{Gutowski:2010}. 
In addition, two special cases are considered in detail. In the first, we consider solutions
for which the  Gauduchon-Tod base space is the Berger sphere. In the second case we take the
null 1-form Killing spinor bilinear to be recurrent, so that the holonomy of the Levi-Civita connection is 
contained in $Sim(3)$. 
\end{minipage}
\end{center}

\begin{quote}
\end{quote}
\end{titlepage}
%%%%%%%%%%%%%%%%%%%%%%%%%%%%%%%%%%%%%%%%%%%%%%%%%%%%%%%
\section{Introduction}
Gravitational models with vanishing or negative cosmological constant can be embedded in supergravity theories. In this sense, finding gravitational solutions in supergravity with zero or negative cosmological constant is simplified through looking for configurations admitting supersymmetry (see \emph{e.g.} \cite{Gauntlett:2002nw,Gauntlett:2003fk}, respectively). This is because the Killing spinor equations are first order, and their integrability conditions ensure that many of the second order field equations are satisfied. This was used for example to show that $AdS_5$ space is the unique maximally supersymmetric ground state of five-dimensional minimal gauged supergravity, and hence interesting in the context of the AdS/CFT correspondence.

Likewise, backgrounds of positive cosmological constant (de Sitter) are of interest due to processes like the higher-dimensional analogue of the black hole anti-evaporation of \cite{Bousso:1997wi}\footnote{Where it was shown that under certain initial conditions (namely, when the perturbation away from the solution is non-vanishing, but its time derivative is) a four-dimensional quantum, asymptotically-de Sitter, Schwarzschild solution of nearly-maximal mass would grow, contrary to a classical thermodynamical intuition.}, or in the study of the proposed dS/CFT correspondence \cite{Strominger:2001pn}, where bulk quantum gravity (or a Supergravity approximation) on a $d$-dimensional de Sitter space would be dual to a Euclidean CFT on $S^{d-1}$ (see also \cite{Hull:1998vg,Hull:1999mt,Liu:2003qaa}). Unfortunately, finding solutions with positive cosmological constant is not easy because gravity with a positive cosmological constant is generally\footnote{A notable exception is \cite{Freedman:1976uk}, in which a four-dimensional, locally diffeomorphism-covariant, $\mathcal{N}=1$ supersymmetric and chiral-invariant field theory for the $(2,3/2)$ and $(1,1/2)$ multiplets, where the spin-1 gauge field is coupled to the fermions, is presented.}
not compatible with supersymmetry \cite{Pilch:1984aw,Lukierski:1984it}. However, one may consider a fake supergravity, obtained from the genuine one via analytic continuation. These kind of constructions were originally proposed in \cite{Freedman:2003ax} to study the stability of domain-wall solutions (see also \cite{Celi:2004st}), and later used to suggest the correspondence between fake supersymmetric domain-wall solutions and pseudo-supersymmetric cosmological solutions \cite{Skenderis:2007sm,Bergshoeff:2007cg}. Contrary to the latter, we are not interested in preserving the original supersymmetry, and we consider fake supergravity as a method for generating cosmological solutions by solving the (pseudo)-Killing spinor equations arising from the analytically continued supersymmetry transformations of the fermionic fields in the original theory. In this regard, our solutions will generally not be supersymmetric, only achieving that in the limiting case of vanishing cosmological constant. The first analysis of pseudo-supersymmetric five dimensional de Sitter solutions in this setting was  performed in \cite{klemm, behrndtcvetic}. The first systematic analysis, not depending on choosing a particular ansatz for the solution,  was performed in \cite{guthktd5,Grover:2009ze}. There, the ``timelike" solutions (those for which the 1-form Killing spinor bilinear is timelike) of the $d=5$ minimal theory were given in terms of a four-dimensional hyper-K\"{a}hler with torsion (HKT) base space. Similarly, $d=5$ fakeSUGRA coupled to abelian vectors was studied in \cite{jansabra}.

Following from this, in \cite{gtnull5d}, solutions of the minimal five dimensional de Sitter supergravity with a null Killing spinor 1-form bilinear were shown to be given by a family of backgrounds with Gauduchon-Tod base space. Gauduchon-Tod spaces \cite{gaudtod} are special types of Einstein-Weyl 3-spaces, which were discovered in the context of hyper-hermitian spaces admitting a tri-holomorphic Killing vector field. We also mention that these spaces appear in the classification of solutions in four-dimensional de Sitter supergravity \cite{Gutowski:2009vb, Meessen:2009ma}, also as certain examples of solutions in five-dimensional de Sitter supergravity with hyper-hermitian isometries \cite{jansabra}, as well as in the classification of four-dimensional (Euclidean) gravitational instantons \cite{euclid}.

The solutions of \cite{gtnull5d}, just like the solutions of ungauged \cite{Gauntlett:2002nw} and gauged (anti-de Sitter) \cite{Gauntlett:2003fk} theories, admit a geodesic, expansion-free, twist-free and shear-free null vector field $N.$ These geometries are known in four-dimensional general relativity as \textit{Kundt metrics} \cite{Kundt:1961}. In higher dimensions, Kundt geometries have been considered in \cite{Coley:2009ut,Podolsky:2008ec,Brannlund:2008zf}. In Minkowski or AdS theories, the null vector is always Killing; and for some special cases it becomes covariantly constant, and the Kundt geometries are \textit{pp-waves}. However this turns out not to be the case in dS theory. In contrast the null vector of solutions of \cite{gtnull5d} is not Killing, but has an interesting property that under certain conditions it becomes \textit{recurrent} with respect to the Levi-Civita connection.

Our work generalises\footnote{One can see that the solutions of \cite{gtnull5d} are a limiting case of the ones presented here, where only the gravity supermultiplet was present.} the analysis of \cite{gtnull5d} to de Sitter $D=5$ supergravity coupled to Abelian vector multiplets. To obtain this theory,  we analytically continue the supersymmetry transformations of the gravitini as well as the gauginos. The vanishing of these transformations produces fake Killing spinor equations, and we consider pseudo-supersymmetric solutions admitting non-vanishing (fake) Killing spinors satisfying these equations.

We outline our work as follows. Section two gives a summary of the basic equations, conventions and the (fake) Killing spinor equations of de Sitter supergravity. In section three we analyse the Killing spinor equations, focusing on the case where the 1-form spinor bilinear is null. The conditions obtained from the gravitino equation are derived from the analysis of the minimal pseudo-supersymmetric solutions in \cite{gtnull5d}. We also introduce local co-ordinates, and show that the  solutions are given in terms of a one-parameter family of  three-dimensional Gauduchon-Tod (GT) spaces. Imposing pseudo-supersymmetry, together with the Bianchi identity and the gauge field equations is sufficient to ensure that all the remaining equations, with the exception of one component o the Einstein equations, hold automatically. We find that the solutions generically preserve half of the pseudo-supersymmetry. As is the case with the classification of supersymmetric backgrounds to genuine supergravity theories, the obstruction to having maximally pseudo-supersymmetric solutions is given by the gaugino equation, which is of an algebraic nature, and sets a condition on the pseudo-Killing spinor. We recall a statement from \cite{Meessen:2001vx}, which conjectures that the only theories (with gaugino fields) that admit non-trivial\footnote{Those not present in the minimal version of the theory.} maximally supersymmetric backgrounds are chiral ones. We provide another instance for this phenomenon, as we find that the only maximally pseudo-supersymmetric background is five-dimensional de Sitter space, just like in the minimal theory of \cite{gtnull5d}\footnote{This result was not presented at the time, and we include it here for completeness.}
In section four we give some simple examples of our solutions, including the near-horizon geometries found in \cite{Gutowski:2010}, and an explicit model with one gauge multiplet, which under vanishing of the potential provides non-BPS solutions to a SUGRA theory. In section five we focus on solutions where the GT-space is the Berger sphere \cite{gaudtod}, \cite{Berger:1961}. In section six, we consider the conditions for which the null Killing spinor 1-form bilinear is recurrent, and investigate the properties of various scalar curvature invariants.

\section{De Sitter $N=2$ supergravity and Killing spinors}
\label{sec:fSUGRA}
In this section, we briefly summarise the supergravity theory we shall consider, together with 
some conventions.
The model which we will consider is $N=2$, $D=5$ gauged supergravity coupled 
to abelian vector multiplets \cite{gunaydin} whose bosonic action is given by
\begin{equation}
S={\frac{1}{16\pi G}}\int \left( -R+2g^{2}{\mathcal{V}}\right)
{\mathcal{\ast }}1-Q_{IJ}\left( -dX^{I}\wedge \star dX^{J}+F^{I}\wedge \ast
F^{J}\right) -{\frac{C_{IJK}}{6}}F^{I}\wedge F^{J}\wedge A^{K}
\label{action}
\end{equation}
where $I,J,K$ take values $1,\ldots ,n$ and $F^{I}=dA^{I}$ are the two-forms
representing gauge field strengths (one of the gauge fields corresponds to
the graviphoton). The constants $C_{IJK}$ are symmetric in $\{I,J,K\}$, we will
assume that $Q_{IJ}$ (the gauge coupling matrix) is positive-definite and invertible,
with inverse $Q^{IJ}$. The $X^{I}$ are scalar fields subject to the constraint
\begin{equation}
{\frac{1}{6}}C_{IJK}X^{I}X^{J}X^{K}=X_{I}X^{I}=1\,.  \label{eqn:conda}
\end{equation}
The fields $X^{I}$ can thus be regarded as being functions of $n-1$
unconstrained scalars $\phi ^{r}$. We list some useful relations associated with $N=2$, $D=5$ 
gauged supergravity
\begin{eqnarray}
Q_{IJ} &=&{\frac{9}{2}}X_{I}X_{J}-{\frac{1}{2}}C_{IJK}X^{K}\ ,\nn
\text{}Q_{IJ}X^{J} &=&{\frac{3}{2}}X_{I}\qquad Q_{IJ}dX^{J}= -{\frac{3}{2}}dX_{I}\ , \nn
{\mathcal{V}} &=&9V_{I}V_{J}(X^{I}X^{J}-{\frac{1}{2}}Q^{IJ}) \ ,
\end{eqnarray}
here $V_{I}$ are constants.
The de Sitter supergravity theory is obtained by sending $g^{2}$ to $-g^{2}$ in ({\ref{action}}).

Pseudo-supersymmetric de Sitter
solutions admit a Dirac spinor $\eta$ satisfying a
gravitino and gaugino Killing spinor equation.
The gravitino Killing spinor equation is:
\begin{equation}
\label{grav}
\left[ \nabla _{M}+{\frac{1}{8}}\Gamma _{M}H_{N_{1}N_{2}}\Gamma
^{N_{1}N_{2}}-{\frac{3}{4}}H_{M}{}^{N}\Gamma _{N}-
g(\frac{1}{2}X\Gamma _{M}-\frac{3}{2}A{}_{M})\right] \eta =0, 
\end{equation}
where we have defined
\begin{equation}
V_{I}X^{I}=X,\text{ \ \ \ \ \ }V_{I}A^{I}{}_{M}=A{}_{M},\text{ \ \ \ \ \ \ }
X_{I}F^{I}{}_{MN}=H_{MN}.
\end{equation}
The gaugino Killing spinor equation is given by
\begin{equation}
\label{dkse}
\left( (-F_{MN}^{I}+X^{I}H_{MN})\Gamma ^{MN}+2\nabla _{M}X^{I}\Gamma
^{M}-4gV_{J}(X^{I}X^{J}-{\frac{3}{2}}Q^{IJ})\right) \eta =0 \ .
\end{equation}
We shall adopt a mostly minus signature for the metric, which we write in a null frame as:
\be
ds^2 = 2 \bbe^+ \bbe^- - \delta_{ij} \bbe^i \bbe^j
\ee
for $i,j,k=1,2,3$.

\section{Analysis of gravitino Killing spinor equation}
\label{subsec:KSE}
We proceed with the analysis of the Killing spinor equations, concentrating on the case for which the 1-form spinor bilinear
generated from the Killing spinor is null; our basis is chosen such that this bilinear 1-form is $\bbe^-$.

The analysis of the gravitino Killing spinor equation has already been completed in the case of the minimal de Sitter supergravity theory in \cite{gtnull5d}, using spinorial geometry techniques introduced in \cite{papadopd11}, in which spinors are identified with differential forms, and are simplified using appropriately chosen gauge transformations. The conditions on the geometry and the fluxes obtained from ({\ref{grav}}) can be read off from the results of section 3 of \cite{gtnull5d} which are listed in equations (3.1)-(3.20). It should be noted that we do not incorporate any of the conditions obtained from the Bianchi identity in \cite{gtnull5d} here, because the 2-form flux $H$ which appears in ({\ref{grav}}) is {\it not} the exterior derivative of $A$, in contrast to the minimal theory. Also, in order to establish the correspondence between the gravitino eq at leastuation solved in \cite{gtnull5d} and ({\ref{grav}}) one makes the replacements
\be
\label{replacements}
F \rightarrow {\sqrt{3} \over 2} H, \qquad \chi \rightarrow - 2 \sqrt{3} g X, \qquad \chi A \rightarrow -3g A
\ee
where the quantities on the LHS of these expressions are the field strength, cosmological constant, and
gauge potential of the minimal theory using the conventions of \cite{gtnull5d}.

Following the reasoning set out in
\cite{gtnull5d}, one can without loss of generality work in a gauge for which the conditions on the geometry are
\bea
\label{geo1}
d \bbe^- &=&0
\nn
d \bbe^+ &=& -3g \bbe^+ \wedge A - \omega_{-,-i} \bbe^- \wedge \bbe^i - \omega_{[i,|-| j]} \bbe^i \wedge \bbe^j
\nn
d \bbe^i &=& 2 \omega_{[-,j]}{}^i \bbe^- \wedge \bbe^j + {\cal{B}} \wedge \bbe^i + 3gX \star_3 \bbe^i
\nn
\cL_N \bbe^i &=&0
\eea

\noindent where $N$ is the vector field dual to $\bbe^-$, and $\omega$ is the five-dimensional spin connection.
${\cal{B}}$ is a 1-form given in terms of the spin connection by
\be
{\cal{B}}= -2 \omega_{i,+-} \bbe^i \ .
\ee
In addition, the 1-form $A$ satisfies
\be
\label{flux1}
-3g A = - \omega_{-,+-} \bbe^- + {\cal{B}}
\ee
and the 2-form flux $H$ satisfies
\be
\label{flux2}
H = \star_3 {\cal{B}} + gX \bbe^+ \wedge \bbe^- +{1 \over 3} \bbe^- \wedge \star_3 (\omega_{-,ij} \bbe^i \wedge \bbe^j) \ .
\ee
Here $\star_3$ denotes the Hodge dual taken on with respect to the 1-parameter family of 3-manifolds $E$ equipped with metric
\be
ds_E^2 = \delta_{ij} \bbe^i \bbe^j
\ee
whose volume form $\epsilon^3$ satisfies
\be
\Gamma_{ijk} \eta = (\epsilon^3)_{ijk} \eta \ .
\ee
Furthermore, in this gauge, the spinor $\eta$ can be taken to be constant, and satisfies
\be
\label{eq:breakingofSUSY}
\Gamma_+ \eta =0
\ee
or equivalently
\be
\Gamma_{+-} \eta = \eta \ .
\ee
Eq.~(\ref{eq:breakingofSUSY}) is the reason for the configurations to preserve at least half of the pseudo-supersymmetry. Using the argument of \cite{Gauntlett:2002nw}, we conclude that if more than half of the pseudo-SUSY is preserved, then this is completely unbroken, and the solutions are maximally pseudo-supersymmetric.

These can be obtained by first demanding the vanishing of the gaugino KSE eq.~(\ref{dkse}), which gives that the $X_I$ and $X^I$ scalars are constant and that all the field strengths $F^I$ are proportional to $H$. These conditions are taken into account when analysing the gravitino KSE, eq.~(\ref{grav}), which can be cast as that of the minimal theory upon relabelling of fields. One then has to examine the integrability condition obtained from this gravitino KSE, and demand that all the different terms (in equal powers of gammas) vanish. This will give the additional constrains which determine the form of the solutions. Since the integrability condition for the KSE of the minimal gauged theory was given in \cite{Gauntlett:2003fk}, we simply introduce the aforementioned relabelling of fields and send $g$ to $ig$, to find that maximal pseudo-supersymmetry implies $H=0$. The integrability condition then simplifies to
\begin{equation}
R_{abcd}=-g^2 X^2(g_{ac}\,g_{bd}-g_{ad}\,g_{bc})\ ,
\end{equation}
which says that the Ricci scalar is constant and negative, and hence that the spacetime is locally $dS_5$. Furthermore, this scenario might provide a five-dimensional instance for the generalised attractor mechanism of \cite{Kachru:2011ps}, in which, on top of constant scalars, fluxes and Riemann curvature, also the background vectors are constant. It remains to be seen whether one could find an attractor potential whose extremisation with respect to the scalar fields $X_I$ might provide the same conditions as the cancellation of the gaugino KSE, and thus fix the (constant) values of the fields.

\subsection{Analysis of the gaugino Killing spinor equation}
\label{subsec:gauginoKSE}
The gaugino Killing spinor equation ({\ref{dkse}}) can be rewritten as
\bea
\label{aux1}
\bigg(2 F^I_{+-} -2 F^I_{+i} \Gamma_- \Gamma^i - F^I_{ij} \epsilon^{ij}{}_k \Gamma^k
-2 X^I H_{+-} + 2 X^I H_{+i} \Gamma_- \Gamma^i + X^I H_{ij} \epsilon^{ij}{}_k \Gamma^k
\nn
+2 \nabla_+ X^I \Gamma_- + 2 \nabla_i X^I \Gamma^i -4gXX^I +6g Q^{IJ} V_J \bigg) \eta =0
\eea
where we have made use of the identity
\be
\Gamma^{ij} \eta = (\epsilon^3)^{ij}{}_k \Gamma^k \eta
\ee
and $\Gamma^i = - \delta^{ij} \Gamma_j$.
On acting on ({\ref{aux1}}) with the projectors ${1 \over 2}(1 \pm \Gamma_{+-})$ one obtains two equations of the form
\be
(\alpha + \beta_i \Gamma^i)\eta =0
\ee
for real $\alpha$, $\beta_i$. As $\eta$ is nonzero, the only solution of such an equation is $\alpha=0, \beta_i=0$,
and on evaluating the resulting conditions on the fluxes and scalars obtained from these equations, one finds that
\be
\label{sc1}
\cL_N X^I=0
\ee
and
\bea
\label{flux3}
F^I_{+-}&=&3g(X X^I-Q^{IJ} V_J)
\nn
F^I_{ij} &=& X^I (\star_3 {\cal{B}})_{ij} + \epsilon_{ij}{}^k \nabla_k X^I
\nn
F^I_{+i} &=&0
\eea
where we adopt the convention that $(\epsilon^3)_{ijk}=(\epsilon^3)_{ij}{}^k$, i.e. indices on the volume form are raised with
the metric of signature $(+,+,+)$.

\subsection{Introduction of local co-ordinates}
\label{subsec:adaptation}
As $\bbe^-$ is closed, one can introduce local co-ordinates $u, v, y^\alpha$ for $\alpha=1,2,3$ such that
\be
\bbe^- = du , \qquad N={\partial \over \partial v}, \qquad \bbe^i = e^i{}_\alpha dy^\alpha \ .
\ee
We remark that possible $du$ terms in $\bbe^i$ can, without loss of generality, be removed by using
a gauge transformation which leaves the spinor $\eta$ invariant, as described in \cite{gtnull5d}.
The 3-dimensional dreibein $e^i{}_{\alpha}$ does not depend on $v$, but in general depends on $y^\alpha$ and $u$,
as do the scalars $X^I$.  We begin by determining the $v$-dependence of $\bbe^+$ and $F^I$.

In order to determine the $v$-dependence of $\bbe^+$ note that ({\ref{geo1}}) implies that
\be
\label{aux2}
\cL_N \bbe^+ = -3gA
\ee
and ({\ref{flux1}}) and ({\ref{flux3}}) imply that
\be
\label{aux3}
\cL_N A = 3g (X^2-Q^{IJ} V_I V_J) \bbe^-
\ee
and so on combining these expressions one obtains
\be
\cL_N \cL_N \bbe^+ = -9g^2  (X^2-Q^{IJ} V_I V_J) \bbe^- \ .
\ee
On integrating up, ({\ref{aux3}}), together with ({\ref{flux1}}), implies that
one can take, without loss of generality,
\be
\label{Aflux}
A = 3g (X^2-Q^{IJ} V_I V_J) v du -{1 \over 3g} {\cal{B}} \ .
\ee
Note also that if ${\tilde{d}}$ denotes the exterior derivative restricted to hypersurfaces of constant $u, v$ then
({\ref{geo1}}) implies that
\be
\label{ews}
{\tilde{d}} \bbe^i = {\cal{B}} \wedge \bbe^i +3gX \star_3 \bbe^i \ .
\ee
We remark that this structure is found in the study of a special class of Einstein-Weyl spaces, \cite{gaudtod} (see also \cite{duntod}). This Gauduchon-Tod (GT) structure  was also found in the analysis of the null class of pseudo-supersymmetric solutions
in the minimal theory \cite{gtnull5d}{\footnote{A brief introduction to GT spaces is given in the appendix of \cite{gtnull5d}}}.
\par
Furthermore, ({\ref{sc1}}) and ({\ref{geo1}}) imply that
\be
\cL_N {\cal{B}}=0
\ee
and likewise ({\ref{ews}}) implies that
\be
{\tilde{d}} {\cal{B}} +3g \star_3 (X {\cal{B}} +{\tilde{d}} X ) =0 \ .
\ee
It is then straightforward to integrate up ({\ref{aux2}}) to find
\be
\bbe^+ = dv -{9 \over 2} g^2 (X^2-Q^{IJ} V_I V_J) v^2 du + W du + v {\cal{B}} + \phi_i \bbe^i
\ee
where $W$ is a $v$-independent function, and $\phi = \phi_i \bbe^i$ is a $v$-independent 1-form.

Next, we wish to determine the $v$-dependence of the field strengths $F^I$.
First, note that
\be
\label{aux4}
\cL_N F^I = d (i_N F^I) = {\tilde{d}} (3g(X X^I - Q^{IJ} V_J)) \wedge du
\ee
on making use of the Bianchi identity $dF^I=0$. However, from ({\ref{flux3}}) one also finds that
\be
F^I = 3g (X X^I - Q^{IJ} V_J) \bbe^+ \wedge \bbe^- + \star_3 (X^I {\cal{B}}+ {\tilde{d}} X^I) + \bbe^- \wedge S^I
\ee
where
\be
S^I = S^I{}_j \bbe^j \ .
\ee
On taking the Lie derivative of this expression, and comparing with ({\ref{aux4}}) one finds that
\be
\cL_N S^I = 3g (X X^I - Q^{IJ} V_J) {\cal{B}} -3g {\tilde{d}} (X X^I - Q^{IJ} V_J)
\ee
and hence we obtain
\bea
\label{fluxfin}
F^I &=& 3g (X X^I -Q^{IJ}V_J) (v {\cal{B}} + dv + \phi) \wedge du +
  \star_3 (X^I {\cal{B}}+ {\tilde{d}} X^I)
 \nn
 &+& du \wedge \bigg( 3gv \big(  (X X^I - Q^{IJ} V_J) {\cal{B}} - {\tilde{d}} (X X^I - Q^{IJ} V_J)  \big) + T^I \bigg)
 \eea
 where
 \be
 T^I = T^I{}_j \bbe^j
 \ee
 are $v$-independent 1-forms on $E$.

Having found this expression for the flux, we continue by imposing two consistency conditions. First,
we require that $V_I F^I = dA$, where $F^I$ is given by ({\ref{fluxfin}}) and $A$ is obtained from ({\ref{Aflux}}).
One finds the following condition on the 1-forms $T^I$,
\be
\label{cc1}
V_I T^I = -{1 \over 3g} {\dot{\cal{B}}} +3g (X^2-Q^{IJ} V_I V_J) \phi
\ee
where ${\dot{{\cal{B}}}}=\cL_{{\partial \over \partial u}} {\cal{B}}$. We also impose $X_I F^I = H$, where $H$ is given in
({\ref{flux2}}), and we note that
\be
\omega_{-,ij} = -{1 \over 2} (v {\tilde{d}} {\cal{B}} + {\tilde{d}} \phi - \phi \wedge {\cal{B}})_{ij}
+{1 \over 2} ({\dot{\bbe}}^i)_j - {1 \over 2} ({\dot{\bbe}}^j)_i \ .
\ee
One finds an additional condition on $T^I$:
\be
\label{cc2}
X_I T^I = -{1 \over 3} \star_3 \bigg( {\tilde{d}} \phi - \phi \wedge {\cal{B}} + \delta_{ij} {\dot{\bbe}}^i \wedge \bbe^j \bigg) \ .
\ee
Next, we impose the Bianchi identities $dF^I=0$,  one obtains the conditions
\bea
\label{cc3}
{\tilde{d}} T^I = \cL_{{\partial \over \partial u}} \star_3 (X^I {\cal{B}}+ {\tilde{d}} X^I)
+3g {\tilde{d}}(X X^I - Q^{IJ} V_J) \wedge \phi +3g (X X^I-Q^{IJ} V_J) {\tilde{d}} \phi
\nn
\eea
and
\bea
\label{cc4}
\tilde{d} \star_3 \big( X^I {\cal{B}} + {\tilde{d}} X^I \big) =0 \ .
\eea
These conditions  exhaust the content of the Killing spinor equations.

\subsection{The gauge and Einstein equations}
\label{subsec:EOMs}
In addition to pseudo-supersymmetry, we require that the field equations be satisfied. 
We start by evaluating the gauge field equations
\be
\label{geq}
d \star (Q_{IJ} F^J) +{1 \over 4} C_{IJK} F^J \wedge F^K =0
\ee
where the 5-dimensional volume form $\epsilon^5$ is related to the 3-dimensional volume form $\epsilon^3$ by
\be
\epsilon^5 = \bbe^+ \wedge \bbe^- \wedge \epsilon^3 \ .
\ee

One obtains the following condition
\bea
{\tilde{d}} \star_3 (Q_{IJ} T^J) &=& -3g \cL_{{\partial \over \partial u}} \big(({3 \over 2} X X_I - V_I) \epsilon^3 \big)
+{3 \over 2} \phi \wedge {\cal{B}} \wedge {\tilde{d}} X_I
+{3 \over 2} {\tilde{d}} \phi \wedge ({\cal{B}} X_I - {\tilde{d}} X_I)
\nn
&+& 3g \phi \wedge \star_3 \big( ({3 \over 2} X X_I - V_I){\cal{B}} -{3 \over 2} X_I {\tilde{d}} X
+{3 \over 2} X {\tilde{d}}X_I + Q_{IJ} {\tilde{d}} Q^{JN} V_N \big)
\nn
&+&{1 \over 2} C_{IJK} T^J \wedge \star_3 (X^K {\cal{B}}+{\tilde{d}} X^K) \ .
\eea
If one considers the limit of this equation in the pure supergravity case (\emph{i.e.} having only the gravity supermultiplet), it becomes satisfied automatically. This is the rather peculiar feature of the \emph{null} case of minimal fSUGRA theories, which was noticed before \cite{gtnull5d,Gutowski:2009vb}.  
 
Next, we consider the Einstein equations. It is straightforward to show that the integrability conditions of the
Killing spinor equations imply that all components of the Einstein equations hold automatically, with the exception of the ``$-  -$"
component. The Einstein equations are
\bea
R_{\alpha \beta} + Q_{IJ} F^I_{\alpha \mu} F^J_\beta{}^\mu - Q_{IJ} \nabla_\alpha X^I
\nabla_\beta X^J
\nn
 + g _{\alpha \beta} \bigg(-{1 \over 6} Q_{IJ} F^I_{\beta_1 \beta_2}
F^{J \beta_1 \beta_2}-6g^2 ({1 \over 2} Q^{IJ}-X^I X^J) V_I V_J \bigg) =0
\eea
and hence the ``$-  -$" component is
\be
R_{--} - Q_{IJ} F^I_{-i} F^J_{-j} \delta^{ij} - Q_{IJ} \nabla_- X^I \nabla_- X^J=0 \ .
\ee
This equation imposes the additional condition
\bea
{\tilde{\nabla}}^2 W + {\tilde{\nabla}}^i (W {\cal{B}}_i) - {\tilde{\nabla}}^i {\dot{\phi}}_i
-3g \phi^i V_I (T^I)_i
- (\ddot{{\bf{e}}}^i)_i -3 (\dot{{\bf{e}}}^j)_i X_I (\star_3 T^I)^i{}_j
\nn
+{1 \over 2} C_{IJK} X^K \big( (T^I)_i (T^J)^i + {\dot{X}}^I \dot{X}^J \big) =0 \ .
\eea
We also require that the solution satisfy the scalar field equations.
However, the integrability conditions of the Killing spinor equations, together with the gauge field equations, imply that
the scalar field equations hold with no additional conditions.

\subsection{Summary}
\label{subsec:summary}
In order to construct a supersymmetric solution in the null class, we introduce local co-ordinates $u, v, y^\alpha$, together with a family of Gauduchon-Tod 3-manifolds ${\rm GT}$, and write the metric as
\be
\label{Kundtmetric}
ds^2 = 2 \bbe^+ \bbe^- - ds_{{\rm GT}}^2, \qquad ds_{{\rm GT}}^2 = \delta_{ij} \bbe^i \bbe^j, \qquad \bbe^i = {e_\alpha}^i dy^\alpha
\ee
where the basis elements $\bbe^i$ do not depend on $v$, but can depend on $u$, and
\bea
\bbe^- &=& du
\nn
\bbe^+ &=& dv -{9 \over 2} g^2 v^2 (X^2-Q^{IJ} V_I V_J) du +  W du + v {\cal{B}} + \phi_i \bbe^i
\eea
where $W$ is a $v$-independent function, and ${\cal{B}}= {\cal{B}}_i \bbe^i$, $\phi = \phi_i \bbe^i$ are $v$-independent 1-forms. The metric (\ref{Kundtmetric}) is actually a Kundt wave, of which we will say more in section \ref{sec:properties}. Because the base-space is Gauduchon-Tod, the basis elements $\bbe^i$ satisfy
\be
\label{GT}
{\tilde{d}} \bbe^i = {\cal{B}} \wedge \bbe^i +3gX \star_3 \bbe^i
\ee
where ${\tilde{d}}$ is the exterior derivative restricted to hypersurfaces of constant $v, u$
and $\star_3$ is the Hodge dual on ${\rm GT}$. The scalars satisfy
\bea
\label{cond1}
\tilde{d} \star_3 \big( X^I {\cal{B}} + {\tilde{d}} X^I \big) &=&0
\eea
and
\bea
\label{cond2}
{\tilde{d}} {\cal{B}} +3g \star_3 (X {\cal{B}} +{\tilde{d}} X ) &=&0 \ .
\eea
The field strengths are
\bea
\label{fs}
F^I &=& 3g (X X^I -Q^{IJ}V_J) ( dv + \phi) \wedge du +
  \star_3 (X^I {\cal{B}}+ {\tilde{d}} X^I)
 \nn
 &&+ du \wedge \bigg(- 3gv{\tilde{d}} (X X^I - Q^{IJ} V_J)  + T^I \bigg)
 \eea
 where
 \be
 T^I = T^I{}_j \bbe^j
 \ee
 are $v$-independent 1-forms on ${\rm GT}$. The 1-forms $T^I$ must further satisfy
 \bea
\label{gcond}
{\tilde{d}} \star_3 (Q_{IJ} T^J) &=& -3g \cL_{{\partial \over \partial u}} \big(({3 \over 2} X X_I - V_I) {\rm dvol}_{{\rm GT}} \big)
+{3 \over 2} \phi \wedge {\cal{B}} \wedge {\tilde{d}} X_I
\nn
&+&{3 \over 2} {\tilde{d}} \phi \wedge ({\cal{B}} X_I - {\tilde{d}} X_I)
+{1 \over 2} C_{IJK} T^J \wedge \star_3 (X^K {\cal{B}}+{\tilde{d}} X^K)
\nn
&+& 3g \phi \wedge \star_3 \big( ({3 \over 2} X X_I - V_I){\cal{B}} -{3 \over 2} X_I {\tilde{d}} X
+{3 \over 2} X {\tilde{d}}X_I + Q_{IJ} {\tilde{d}} Q^{JN} V_N \big)
\nn
\eea
and
\be
\label{VT}
V_I T^I = -{1 \over 3g} {\dot{\cal{B}}} +3g (X^2-Q^{IJ} V_I V_J) \phi
\ee
and
\be
\label{XT}
X_I T^I = -{1 \over 3} \star_3 \bigg( {\tilde{d}} \phi - \phi \wedge {\cal{B}} + \delta_{ij} {\dot{\bbe}}^i \wedge \bbe^j \bigg)
\ee
and
\be
\label{dT}
{\tilde{d}} T^I = \cL_{{\partial \over \partial u}} \star_3 (X^I {\cal{B}}+ {\tilde{d}} X^I)
+3g {\tilde{d}} \bigg( (X X^I - Q^{IJ} V_J) \phi \bigg) \ .
\nn
\ee
Finally, the function $W$ is found by solving
\bea
\label{W}
{\tilde{\nabla}}^2 W + {\tilde{\nabla}}^i (W {\cal{B}}_i) - {\tilde{\nabla}}^i {\dot{\phi}}_i
-3g \phi^i V_I (T^I)_i
- (\ddot{{\bf{e}}}^i)_i -3 (\dot{{\bf{e}}}^j)_i X_I (\star_3 T^I)^i{}_j
\nn
+{1 \over 2} C_{IJK} X^K \big( (T^I)_i (T^J)^i + {\dot{X}}^I \dot{X}^J \big) =0 \ .
\eea
We remark that ({\ref{gcond}}) and ({\ref{dT}}) and ({\ref{W}}) always admit solutions, however it is not a priori apparent that
({\ref{VT}}) and ({\ref{XT}}) can always be solved.

\section{Some simple examples}
\label{sec:examples}
\subsection{Near-horizon geometries}
\label{subsec:nearhorizon}

The near-horizon geometries found in  \cite{Gutowski:2010}
are all examples of pseudo-supersymmetric solutions in the null class.
We take $\frac{\partial}{\partial u}$ as a symmetry of the full solution, and set the  $X^I$ to be constant, and
 $W=0$, $\phi =0$, $T^I=0$. Then the remaining conditions on the geometry simplify considerably, and one finds
\bea
\bbe^- &=& du
\nn
\bbe^+ &=& dv -{9 \over 2} g^2 (X^2-Q^{IJ} V_I V_J) v^2 du + v {\cal{B}}
\nn
\bbe^{\,i\,} &=& {e_\alpha}^i\, dy^\alpha
\eea
where
\bea
{\tilde{d}} \bbe^i = {\cal{B}} \wedge \bbe^i +3gX \star_3 \bbe^i
\eea
and the gauge field strengths are
\bea
F^I = 3g (X X^I -Q^{IJ} V_J)\,  dv \wedge du + X^I \star_3 {\cal{B}}\ ,
\eea
with the 1-form ${\cal{B}}$ satisfied
\bea
{\tilde{d}} {\cal{B}} + 3gX \star_3 {\cal{B}}=0\ .
\eea
This solution can be interpreted as the (pseudo-supersymmetric) near-horizon geometry of a (possibly non-pseudo-supersymmetric) black hole. The case for which  $V_IX^I=0,\, {\cal{B}}\neq 0$  is of particular interest, as the spacetime geometry
is $M_3 \times S^2$, where $M_3$ is a $U(1)$ vibration over $AdS_2$ related to the near-horizon extremal Kerr solution,
and the spatial cross-sections of the event horizon are $S^1 \times S^2$.

\subsection{A small model of Real Special Geometry}
\label{sec:simplemodel}
A particularly simple class of solutions  can be constructed with only one vector multiplet. Since the $5$-dimensional gravity multiplet does not contain scalars, there is only one physical scalar field $\varphi$, and we choose the only non-vanishing value for the symmetric constant $C_{IJK}$ to be given by \mbox{$C_{122}=1$}. With this choice,

\begin{displaymath}
X^I=\left(\begin{tabular}{c}$\varphi^{-2}$\\
$\sqrt{2}\varphi$\end{tabular}\right),\
X_I=\frac{1}{3}\left(\begin{tabular}{c}$\varphi^2$\\
$\sqrt{2}\varphi^{-1}$\end{tabular}\right),\ 
\end{displaymath}
\begin{equation}
\label{eq:modelconstraints}
Q_{IJ}=\frac{1}{2}\text{diag}\,(\varphi^4,\varphi^{-2}),\quad Q^{IJ}=2\text{diag}\,(\varphi^{-4},\varphi^{2})\ .
\end{equation}
The equations resulting from the classification of the theory, summarised in subsection \ref{subsec:summary}
must also be satisfied, however we shall not present this analysis here.

This model is interesting as it provides a simple setting for non-supersymmetric solutions to a supersymmetric theory. The potential is given by
\begin{equation}
{\mathcal{V}}=9V_2(V_2 \varphi^2+2\sqrt{2}V_1 \varphi^{-1})\ .
\end{equation}
One can immediately see if $V_2=0$ the theory is supersymmetric ({\em i.e.} it was the definite-positiveness of $\mathcal{V}$ that granted us a de Sitter-like fSUGRA structure). This solution, however, is non-BPS, since the presence of $V_1$ means its Killing spinor equation is not that of standard five-dimensional SUGRA. This kind of behaviour was already present in the classification of four-dimensional fSUGRA (see section 4 of \cite{Meessen:2009ma}) and one can also here make use of the oxidation/dimensional-reduction relations between supergravity theories (see {\em e.g.} \cite{Lozanoetal:2002} or Section 5.3 in \cite{Gutowskietal:2003}) to obtain solutions to minimal $N=(2,0)$ $d=6$ supergravity.

To achieve this, we make use of the results developed in \cite{Lozanoetal:2002}, which provide the $5$-dimensional action (obtained through a KK compactification over an $S^1$) and compare it to the action for our model. The fields in these actions, however, do not promptly correspond, and they have to be appropriately identified. The dimensionally-reduced action is given by\footnote{We have adapted the conventions of \cite{Lozanoetal:2002} to those of here. In particular, the Riemmann tensor has the opposite defining sign.}
\begin{equation}
\label{reducedaction}
S=\int d^5 x\, \sqrt{|g|}\,k \left( -R-\frac{1}{4}k^2F^2(A)-\frac{1}{4}k^{-2}F^2(B)+\frac{\epsilon^{\mu\nu\rho\sigma\tau}}{8\sqrt{|g|}}k^{-1}F(A)_{\mu\nu} F(B)_{\rho\sigma} B_\tau\right)\ .
\end{equation}
The kinetic term for the graviton in this action does not have a canonical form, so we proceed to rescale the metric by a scalar $k$, and hence unveil the kinetic term for the scalars hidden in the Einstein-Hilbert term
\begin{equation}
\label{metricrescaling}
g_{\mu\nu} \rightarrow k^{\frac{-2}{3}} g_{\mu\nu}\ .
\end{equation}
%This entails that
%\be
%\begin{tabular}{c}
%$\sqrt{|g|} \rightarrow k^{\frac{-5}{3}}\sqrt{|g|}\ ,\qquad R\rightarrow k^{\frac{2}{3}}\left( R-12(\partial\: ln\: k^{\frac{-1}{3}})^2-8\nabla^2\:ln\: k^{\frac{-1}{3}}\right)\ ,$\\
%$F^2 \rightarrow k^\frac{4}{3} F^2\ .$
%\end{tabular}
%\ee
The action hence becomes
\begin{equation} 
\label{eq:scaledaction}
S=\int d^5x\, \sqrt{|g|} \left( -R+\frac{4}{3}k^{-2} (\partial k)^2 -\frac{1}{4}k^{\frac{8}{3}} F^2(A)-\frac{1}{4}k^{-\frac{4}{3}} F^2(B)+\frac{\epsilon}{8\sqrt{|g|}} F(A)F(B)B \right)\ .
\end{equation}

We then compare this action with that of our model
\bea
\label{actionmodel}
S&=&\int  d\text{vol}\left( -R+3\varphi^{-2}(\partial\varphi)^2-\frac{1}{4}\varphi^{4}(F^{1})^2- \frac{1}{4}\varphi^{-2} (F^{2})^2\right.\nn
&&\hspace{4em}\left. -\frac{\epsilon}{24\sqrt{|g|}} F^{2}F^{2}A^{1}-\frac{\epsilon}{12\sqrt{|g|}} F^{1}F^{2}A^{2}\right)\ ,
\eea
where the topological term is integrated by parts to identify the gauge fields. Upon inspection, $k=a\varphi^{\frac{3}{2}}$, where $a$ is just a real constant of integration, and
\be
\begin{tabular}{cc}
$A=-a^{-\frac{4}{3}} A^1$\\
$B=\pm a^{\frac{2}{3}} A^2\ .$
\end{tabular}
\ee
To identify the remaining elements we consider the supersymmetric variation of the gaugino, \emph{i.e.} eq.~(\ref{dkse}), and that obtained from the dimensional reduction of the gravitino KSE of the $6$-dimensional theory (see \emph{e.g.} eq.~(1.15) in \cite{Lozanoetal:2002}). We obtain that $A_\mu=-A^1_\mu$, $B_\mu=-A^2_\mu$ and $k=-\varphi^{\frac{3}{2}}$.

This gives the identification of fields, and one can use the equations of the reduction on a circle to obtain the six-dimensional ones\footnote{In flat indices, the six-dimensional space is labelled by $a=\{0,1,2,3,4\}$ and $\sharp$, where $a$ span the five-dimensional space.} 
\be
\label{eq:6dsolutions}
\begin{tabular}{c}
$g^{(6)}_{\mu\nu}=g^{(5)}_{\mu\nu}- \varphi^3 A^1_\mu A^1_\nu\ ,\hspace{1cm}g^{(6)}_{\mu\sharp}=-\varphi^3 A^1_\mu\ ,\hspace{1cm}g^{(6)}_{\sharp \sharp}=-\varphi^3\ ,$\\
\\
$H^-_{ab\sharp}=\varphi^{-\frac{3}{2}} F^2_{ab}\ ,\hspace{1cm} H^-_{abc}=-\varphi^{-\frac{3}{2}}\left(\star_\text{(6)}(e^\sharp \wedge F^2)\right)_{abc}\ .$\\
\end{tabular}
\eea
The geometries obtained in this manner, lifted from solutions to the model of (\ref{eq:modelconstraints}) that fulfill the constraining equations of subsection \ref{subsec:summary}, are solutions to minimal $N=(2,0)$ $d=6$ SUGRA with (the bosonic part of the) action given by
\be
\label{6daction}
\int d^6x \sqrt{|g|} \left( R+\frac{1}{24}(H^-)^2\right)\ ,
\ee
where as usual one considers the anti-self-duality of $H^-$ as an additional constraint on the theory, rather than on the actual action.

\section{Solutions with compact Einstein-Weyl structure}
\label{subsec:Berger}

In this section, we concentrate on the case for which the GT-space is compact and without boundary for all $u$, and
$X^I, T^I, \phi, W$ are smooth. The near-horizon geometries of the previous section are special examples of such solutions.

To proceed, consider ({\ref{cond1}}) and contract with $X^I$. One finds that
\bea
{\tilde{\nabla}}^i {\cal{B}}_i +{2 \over 3} Q_{IJ} {\tilde{\nabla}}^i X^I {\tilde{\nabla}}_i X^J =0 \ .
\eea
On integrating over ${\rm GT}$ one finds
\bea
\int_{{\rm GT}} Q_{IJ} {\tilde{\nabla}}^i X^I {\tilde{\nabla}}_i X^J =0
\eea
and assuming that $Q_{IJ}$ is positive definite, this implies that $X^I = X^I(u)$, $X=X(u)$.

We shall consider ({\ref{GT}}) with $X \neq 0$ and take ${\rm GT}$ to be the Berger sphere{\footnote{If $X=0$ then ${\rm GT}$ is either
$S^1 \times S^2$ or $T^3$, according as ${\cal{B}} \neq 0$ or ${\cal{B}}=0$ respectively. We do not consider these cases here.}} \cite{gaudtod}, \cite{Berger:1961}; one can write
\bea
\label{berg1}
ds^2_{{\rm GT}}&=&  {\cos^2 \mu \over 9 g^2 X^2} \bigg( \cos^2 \mu (\sigma^3_L)^2  + (\sigma^1_L)^2+(\sigma^2_L)^2 \bigg)
\nn
{\cal{B}} &=& \sin \mu \cos \mu \sigma^3_L
\eea
where $\mu=\mu(u)$, and $\sigma^i_L$ are the left-invariant 1-forms on $SU(2)$ satisfying
 \[\tilde{d}\sigma_L^i=-\frac{1}{2}\epsilon^{ijk}\,\sigma_L^j\wedge \sigma_L^k\ .\]

Next, note that ({\ref{dT}}), ({\ref{VT}}) can be solved, making use of ({\ref{cond2}}), to give
\bea
\label{texp}
T^I = - {\cal{L}}_{\partial \over \partial u} \bigg( {X^I \over 3gX} {\cal{B}} \bigg)
+3g (X X^I - Q^{IJ} V_J ) \phi + \Theta^I
\eea
where $\Theta^I$ are 1-forms on ${\rm GT}$ satisfying
\bea
\label{tq1}
V_I \Theta^I =0
\eea
and
\bea
\label{tq2}
{\tilde{d}} \Theta^I =0 \ .
\eea

Next, simplify ({\ref{gcond}}) using ({\ref{XT}}) and ({\ref{dT}}), using the identity
\bea
{\tilde{d}} \bigg(  \delta_{ij} {\dot{\bbe}}^i \wedge \bbe^j  \bigg) =2 {\cal{B}} \wedge  \bigg(  \delta_{ij} {\dot{\bbe}}^i \wedge \bbe^j  \bigg)
+9 g {\dot{X}} {\rm dvol_{GT}} +3 gX {\cal{L}}_{\partial \over \partial u} {\rm dvol_{GT}} \ .
\eea 

After some manipulation, one finds that ({\ref{gcond}}) can be rewritten as
\bea
{\tilde{d}} \bigg( {3 \over 4} X_I   \delta_{ij} {\dot{\bbe}}^i \wedge \bbe^j  
-{1 \over 3gX} Q_{IJ} T^J \wedge {\cal{B}} -{1 \over X} V_I \phi \wedge {\cal{B}} \bigg)
\nn
+{9 \over 2} g \big({1 \over 2}  {\dot{X}} X_I -X {\dot{X}}_I \big) {\rm dvol_{GT}}
+ \big(3g V_I -{9 \over 4} g X X_I \big) {\cal{L}}_{\partial \over \partial u} {\rm dvol_{GT}}
\nn
+{1 \over 2gX} \bigg( - {\dot{X}}_I {\star_3 {\cal{B}}} \wedge {\cal{B}} + X_I ({\cal{L}}_{\partial \over \partial u} 
\star_3 {\cal{B}}) \wedge {\cal{B}} \bigg) = 0 \ .
\eea
For the Berger sphere, the second and third lines of this expression can be written in the form $Q^I(u) {\rm dvol_{GT}}$,
and hence on integrating over ${\rm GT}$ one obtains two separate conditions:
\bea
\label{ccx1}
{\tilde{d}} \bigg( {3 \over 4} X_I   \delta_{ij} {\dot{\bbe}}^i \wedge \bbe^j  
-{1 \over 3gX} Q_{IJ} T^J \wedge {\cal{B}} -{1 \over X} V_I \phi \wedge {\cal{B}} \bigg) =0
\eea
and
\bea
\label{ccx2}
{9 \over 2} g \big({1 \over 2}  {\dot{X}} X_I - X {\dot{X}}_I \big) {\rm dvol_{GT}}
+ \big(3g V_I -{9 \over 4} g X X_I \big) {\cal{L}}_{\partial \over \partial u} {\rm dvol_{GT}}
\nn
+{1 \over 2gX} \bigg( - {\dot{X}}_I {\star_3 {\cal{B}}} \wedge {\cal{B}} + X_I ({\cal{L}}_{\partial \over \partial u} 
\star_3 {\cal{B}}) \wedge {\cal{B}} \bigg) = 0 \ .
\eea
On using ({\ref{texp}}), ({\ref{ccx1}}) and ({\ref{XT}}) can be rewritten as
\bea
\label{ddx1}
{\tilde{d}} \bigg( \star_3 \Theta^I +{1 \over 3gX} \Theta^I \wedge {\cal{B}}
+3g (X X^I - Q^{IJ} V_J) \star_3 \phi +{1 \over 2} X^I   \delta_{ij} {\dot{\bbe}}^i \wedge \bbe^j  \bigg) =0
\eea
and
\bea
\label{XT2}
-{\cal{L}}_{\partial \over \partial u} \bigg( {1 \over 3gX} {\cal{B}} \bigg) +gX \phi + X_I \Theta^I
= -{1 \over 3} \star_3 \bigg({\tilde{d}} \phi - \phi \wedge {\cal{B}} +   \delta_{ij} {\dot{\bbe}}^i \wedge \bbe^j  \bigg) \ .
\eea
On contracting ({\ref{ddx1}}) with $X_I$ and using ({\ref{XT2}}) one finds
\bea
{\cal{L}}_{\partial \over \partial u} {\rm dvol_{GT}} = -3{{\dot{X}} \over X} {\rm dvol_{GT}}
\eea
which for the Berger sphere implies that the squashing of the $S^3$ is $u$-independent, i.e. $\mu$ is constant in 
({\ref{berg1}}), and the
$u$ dependence of the metric on ${\rm GT}$ is in the overall conformal factor of $X^{-2}$. It follows that
\bea
{\cal{L}}_{\partial \over \partial u} \star_3 {\cal{B}} = -{{\dot{X}} \over X} \star_3 {\cal{B}}
\eea
and hence ({\ref{ccx2}}) can be simplified to give
\bea
\label{ccx3}
{\dot{X}} X_I -{1 \over 2} X {\dot{X}}_I - {{\dot{X}} \over X} V_I
+{1 \over 18g^2X^2} \big(- X {\dot{X}}_I - {\dot{X}} X_I \big) {\cal{B}}^2 =0 \ .
\eea 
On contracting this expression with $X^I$ and $Q^{IJ} V_J$
one finds
\bea
{\dot{X}} {\cal{B}}^2 =0, \qquad (X^2-Q^{IJ} V_I V_J) {\dot{X}} =0 \ .
\eea

Suppose first that $X^2- Q^{IJ} V_I V_J \neq 0$. Then ${\dot{X}}=0$ and ({\ref{ccx3}}) further implies that 
\bea
{\dot{X}}_I =0
\eea
also. Furthermore, contracting ({\ref{ddx1}}) with $V_I$ gives
\bea
{\tilde{d}} \star_3 \phi =0
\eea
so ({\ref{ddx1}}) is
\bea
{\tilde{d}} \star_3 \Theta^I + \Theta^I \wedge \star_3 {\cal{B}}=0 \ .
\eea
As ${\tilde{d}} \Theta^I=0$ and the Berger sphere is simply connected, this implies that the $\Theta^I$ are exact
\be
\Theta^I = {\tilde{d}} H^I
\ee
where
\bea
{\tilde{\nabla}}^2 H^I + {\cal{B}}^i {\tilde{\nabla}}_i H^I =0 \ .
\eea
It follows that ${\tilde{d}} H^I=0$, and so $\Theta^I =0$. Also, ({\ref{W}}) implies that $W=W(u)$.
It remains to consider the condition
\bea
\label{auxcc}
gX \phi = -{1 \over 3} \star_3 \big( {\tilde{d}} \phi - \phi \wedge {\cal{B}} \big) \ .
\eea
To proceed, note that the Ricci tensor of ${\rm GT}$ is 
\bea
{\tilde{R}}_{ij} = \bigg( {\cal{B}}^\ell {\cal{B}}_\ell +{9 \over 2} g^2 X^2 \bigg) \delta_{ij} - {\cal{B}}_i {\cal{B}}_j
\eea
as a consequence of ({\ref{GT}}). After some manipulation, it also can be shown that ({\ref{auxcc}})
implies
\bea
{\tilde{\nabla}}^2 \phi^2 + {\cal{B}}^i {\tilde{\nabla}}_i \phi^2 = 2 {\tilde{\nabla}}^{(i} \phi^{j)}
{\tilde{\nabla}}_{(i} \phi_{j)} + 3 \bigg( {\cal{B}}^2 \phi^2 - ({\cal{B}} . \phi)^2 \bigg) \ .
\eea
As the RHS of this expression is a sum of two non-negative terms, it follows from the maximum
principle that $\phi^2$ is constant, and
\bea
{\tilde{\nabla}}_{(i} \phi_{j)} =0, \qquad  {\cal{B}}^2 \phi^2 - ({\cal{B}} . \phi)^2 =0 \ .
\eea
These conditions also imply that one can take, without loss of generality,
\bea
\phi = k \sigma^3_L
\eea
for constant $k$, irrespective of whether ${\cal{B}}$ vanishes or not.
Also note that by making a co-ordinate transformation of the form
\bea
{\hat{u}}= f(u), \qquad v = h(u) {\hat{v}} + g(u), \qquad \psi = {\hat{\psi}} + \ell(u)
\eea
where we take the vector field dual to $\sigma^3_L$ to be ${\partial \over \partial \psi}$,
one can choose the functions $f, h, g, \ell$ appropriately such that the form of the metric and gauge
field strengths is preserved, and in the new co-ordinates, $W=0$.

To summarise; if $X^2-Q^{IJ} V_I V_J \neq 0$, then the Berger sphere squashing
parameter $\mu$ and $X^I$ are constant, and 
\bea
ds^2 &=& 2 du (dv -{9 \over 2} g^2 v^2 (X^2-Q^{IJ} V_I V_J) du + (v \sin \mu \cos \mu + k) \sigma^3_L) 
\nn
&-& {\cos^2 \mu \over 9 g^2 X^2} \bigg( \cos^2 \mu (\sigma^3_L)^2  + (\sigma^1_L)^2+(\sigma^2_L)^2 \bigg)
\nn
F^I &=& 3g (X X^I -Q^{IJ}V_J)  dv \wedge du + {X^I \over 3gX} \sin \mu \cos \mu \  \sigma^1_L \wedge \sigma^2_L \ ,
 \eea
where $k$ is constant.

In the special case $X^2-Q^{IJ} V_I V_J=0$ then there are two possibilities. If ${\cal{B}} \neq 0$ then 
({\ref{ccx3}}) implies that the $X^I$ are again constant, whereas if ${\cal{B}}=0$ then ({\ref{ccx3}}) implies that
\bea
X_I = {2 \over 3} X^{-1} V_I + X^2 Z_I
\eea
for constant $Z_I$. Also ({\ref{ddx1}}), ({\ref{XT2}}) imply
\bea
H^I -{3 \over X} (X X^I-Q^{IJ}V_J) X_N H^N &=& L^I
\eea
and
\bea
\phi = -{1 \over gX} \bigg( {\tilde{d}} (X_I H^I) + X_I H^I {\cal{B}} \bigg) + k(u) \sigma^3_L
\eea
where $H^I$ are functions, $\Theta^I ={\tilde{d}} H^I$ and $L^I=L^I(u)$ satisfy $X_I L^I =0$.

%
%\subsection{A Calderbank-Tod Wave}
%A slighly more involved solution for the theory is given by considering the Gauduchon-Tod space presented in \cite{Calderbank:2001}\\
%\[ds^2=2du(dv -{9 \over 2} g^2 v^2 (X^2-Q^{IJ} V_I V_J) du + W du + v {\cal{B}} + \phi_i\,\bbe^i)-dx^2+|x+h|^2\:ds^2_{S^2}\]
%\begin{equation}
%\label{CTwave} B=\frac{2x+h+\bar{h}}{|x+h|^2}\,dx,\qquad X^I=\frac{i(\bar{h}-h)\,C^I}{3g|x+h|^2}\ ,
%\end{equation}\\
%where $C^I$ is a constant such that $V_I\,C^I=1$, and $h=h(z)$ is a holomorphic function. Since the C\&T wave is a GT space, eqs.~(\ref{eq:cond1}) and (\ref{eq:cond2}) are fulfilled. However, the non-constancy of the $X^I$ spoils the identities (\ref{eq:cond0a}), (\ref{eq:cond0b}) and hence this space is \underline{not} a good cross-section for a Kundt wave background to the fSUGRA theory. 
%

\section{Solutions with a recurrent vector field}
\label{sec:properties}
The geometry which we have found (\ref{Kundtmetric}) is a 5-dimensional Kundt wave \cite{Kundt:1961}. This is a kind of metric in the Walker form \cite{Walker:1950} that admits a null vector generating a geodesic null congruence that is hypersurface orthogonal, non-expanding and shear-free. It is a special case of the higher-dimensional Kundt metrics studied in \cite{Podolsky:2008ec}, \cite{Coley:2009ut}. \par
To see this, consider the null vector field $N=\partial/\partial v$. The congruence of integral curves affinely parametrised by $v$ fulfills $\nabla_N N=0$ (geodesic). $N$ is hypersurface orthogonal, {\em i.e.} $\bbe^- \wedge d \bbe^-=0$,  which means that the congruence is twist-free. It is also non-expanding $\nabla_\mu N^\mu=0$ and shear-free $\nabla_{(\mu} N_{\nu)} \nabla^\mu N^\nu=0$. Hence it is a Kundt metric.

As in the case of minimal de Sitter supergravity, the null vector field $N$ is not Killing. 
We shall consider the necessary and sufficient conditions for $N$ to be recurrent, which places additional
restrictions on the holonomy of the Levi-Civita connection.
Recurrency with respect to this connection is defined as 
\bea
\nabla_\mu N^\nu=C_\mu N^\nu \ ,
\eea
where $C$ is the recurrent one-form \cite{Gibbons:2007}. D-dimensional geometries that allow recurrent vector fields have their holonomy group contained in the Similitude group $Sim(D-2)$. This is a $(D^2-3D+4)/2$-dimensional subgroup of the Lorentz group $SO(D-1,1)$, and it is isomorphic to the Euclidean group $E(D-2)$ augmented by homotheties. It is also the maximal proper subgroup of the Lorentz group, and hence connections admitting $Sim(D-2)$ have a minimal (non-trivial) holonomy reduction.

Theories with the $Similitude$ group have received some attention in the past few years, as they have been shown to hold interesting physical features. They are linked to theories with vanishing quantum corrections \cite{Coleyetal:2008} and to the recently proposed theories of \emph{Very Special Relativity} \cite{Cohen:2006} and \emph{General Very Special Relativity} \cite{Gibbonsetal:2007}. 

For our solutions, note that
\bea
\nabla_- N_j = {1 \over 2} {\cal{B}}_j
\eea
and so a necessary condition for the $N$ to be recurrent is ${\cal{B}}=0$. In fact, it is also sufficient, and one finds that
if ${\cal{B}}=0$ then
\bea
\nabla_\mu N^\nu = -9g^2 (X^2-Q^{IJ} V_I V_J) v N_\mu N^\nu \ .
\eea
In the remainder of this section we take ${\cal{B}}=0$, and investigate the resulting conditions imposed on the geometry,
To proceed, first consider ({\ref{GT}}). If $X \neq 0$ then the ${\rm GT}$ space is $S^3$, whereas if $X=0$ it is flat.
Next consider ({\ref{cond1}}), on contracting with $X_I$, this condition is equivalent to
\bea
Q_{IJ} {\tilde{\nabla}}^i X^I {\tilde{\nabla}}_i X^J =0
\eea
and assuming that $Q_{IJ}$ is positive definite, this implies that $X^I=X^I(u)$.
Then ({\ref{VT}}) and ({\ref{dT}}) further imply that
\bea
T^I = 3g (X X^I - Q^{IJ} V_J) \phi + K^I
\eea
where $K^I$ are 1-forms on ${\rm GT}$ satisfying
\bea
{\tilde{d}} K^I = 0, \qquad V_I K^I =0
\eea
and ({\ref{XT}}) simplifies to
\bea
gX \phi + X_I K^I = -{1 \over 3} \star_3 {\tilde{d}} \phi \ .
\eea

The conditions obtained from ({\ref{gcond}}) are 
\bea
{\dot{X}}=0
\eea
and
\bea
{\tilde{d}} \star_3 K^I = -3g (X X^I-Q^{IJ} V_J) {\tilde{d}} \star_3 \phi +3g X {\dot{X}}^I {\rm dvol_{GT}} \ .
\eea
In particular, note that if $X^2 - Q^{IJ} V_I V_J \neq 0$, then on contracting this condition with $V_I$, one finds that
\bea
{\tilde{d}} \star_3 \phi =0, \qquad {\tilde{d}} \star_3 K^I = 3g X {\dot{X}}^I {\rm dvol_{GT}} \ .
\eea
The function $W$ must satisfy 
\bea
\label{wrecur}
{\tilde{\nabla}}^2 W  & = & {\tilde{\nabla}}^i {\dot{\phi}}_i +9g^2 (X^2 - Q^{IJ}V_I V_J)\, \phi\cdot \phi -{1 \over 2} C_{IJK} X^K \big( (T^I)_i (T^J)^i + {\dot{X}}^I \dot{X}^J \big) \nn
\eea
as a consequence of ({\ref{W}}).

Given $K^I, \phi, X^I, W$ satisfying these conditions, the metric and field strengths are

\bea
\label{Bnullsol}
ds^2 & = & 2 du(dv -{9 \over 2} g^2 v^2 (X^2-Q^{IJ} V_I V_J) du +  W du +  \phi_i \bbe^i)- ds_{{\rm GT}}^2\hspace{1cm}\nn
F^I  & = & 3g (X X^I -Q^{IJ}V_J)\, dv \wedge du - K^I  \wedge du \ .
\eea

As a simple example, take $X \neq 0$, $X^I$ constant with $K^I=0$. Then the Gauduchon-Tod space is $S^3$.
All of the conditions are satisfied if one takes 
$\phi=\xi_i\,\sigma_L^i$, where $\xi_i=\xi_i(u)$,
and with this choice of $\phi$, ({\ref{wrecur}}) implies that $W$ is a ($u$-dependent) harmonic function on $S^3$.
This solution describes gravitational waves propagating through a generalized squashed Nariai universe.
Generically, these waves will be plane-fronted waves, as $N$ is not a Killing vector. However, if $X^2-Q^{IJ}V_I V_J=0$ they are pp-waves.

Alternatively, taking again $X^I$ constant with $X \neq 0$, one can instead set $\phi=0$ and 
\bea
K^I = K^I_i(u) \sigma_L^i \ .
\eea
Then the conditions which must be satisfied are
\bea
V_I K^I= X_I K^I =0, \qquad {\tilde{\nabla}}^2 W = Q_{IJ} (K^I)_i (K^J)^i \ .
\eea
In this case, if $K^I \neq 0$, one must have non-vanishing $W$.

It was shown that for recurrent solutions in the minimal theory, all scalar curvature invariants
constructed purely algebraically from the Riemann tensor are constant \cite{gtnull5d}. By computing the Ricci scalar for 
the recurrent solutions constructed here, it can be seen that a necessary and sufficient condition for the Ricci scalar
to be constant is that $Q^{IJ} V_I V_J$ is constant. This condition is also sufficient to ensure that all
the other algebraic scalar curvature invariants are also constant. To see this, define
\bea
\psi_{\mu \nu \lambda} &=& {2 \over 3} \nabla_\mu \nabla_{[\nu} \phi_{\lambda ]}+
{1 \over 3} \nabla_\nu \nabla_{[\mu} \phi_{\lambda ]} - {1 \over 3} \nabla_\lambda \nabla_{[\mu} \phi_{\nu ]}
\nn
\theta_{\mu \nu} &=& \nabla_\mu \nabla_\nu W +9 g^2(X^2-Q^{IJ} V_I V_J) v \nabla_{(\mu}\phi_{\nu )}
+{1 \over 4} (d \phi)_{\mu \lambda} (d \phi)_{\nu}{}^\lambda \ ,
\eea
and note that
\bea
N^\mu \psi_{\mu \nu \lambda}=0, \qquad N^\mu \theta_{\mu \nu}=0 \ .
\eea
The Riemann tensor satisfies
\bea
R_{\mu \nu \lambda \tau} = (R^0)_{\mu \nu  \lambda \tau}
+4 N_{[\mu} \theta_{\nu ] [ \lambda} N_{ \tau]}
+N_\mu \psi_{\nu \lambda \tau} - N_\nu \psi_{\mu \lambda \tau} + N_\lambda \psi_{\tau \mu \nu}
-N_\tau \psi_{\lambda \mu \nu}
\eea
where $R^0$ is the Riemann tensor of the metric $g^0$ obtained from the metric in ({\ref{Bnullsol}}) setting $W=0, \phi=0$,
where $(R^0)_{\mu \nu \lambda \tau} = g^0_{\mu \kappa} (R^0)^\kappa{}_{\nu \lambda \tau}$. The metric $g^0$
is the metric on $AdS_2 \times {\rm GT}$, $\bR^{1,1} \times {\rm GT}$ and $dS_2 \times {\rm GT}$ according as
$X^2-Q^{IJ} V_I V_J$ is negative, zero or positive. It then follows, from exactly the same reasoning as set out
for the minimal case in \cite{gtnull5d}, that all algebraic scalar curvature invariants constructed from the metric
$g$ and Riemann tensor $R$ are identical to the same invariants constructed from $g^0$, $R^0$ and hence are constant.
The status of scalar curvature invariants constructed from covariant derivatives of the Riemann tensor
remains to be determined.

\par
For ${\cal{B}}\ne 0$, one can also provide the construction with a $Sim$-holonomy structure. This is relevant for the embedding of the Berger sphere considered in section \ref{subsec:Berger}. By considering the gravitino Killing spinor equation, one obtains 
\begin{equation}
\nabla_\mu N_\nu=-3gA_\mu N_\nu + \frac{1}{2}X_I\, [\star (N\wedge F^I)]_{\mu\nu} \ .
\end{equation}
Next, define a new covariant derivative ${\cal{D}}$ by
\begin{equation}
\mathcal{D}_\mu N_\nu\equiv \nabla_\mu N_\nu- {S_{\mu\nu}}^\rho N_\rho= -3gA_\mu N_\nu\ ,\quad \text{where}\ \  S=\frac{1}{2}\star (X_I F^I)
\end{equation}
can be interpreted as a totally antisymmetric torsion 3-form, hence the new connection is metric-compatible.
It is clear that that $N$ is recurrent with respect to the connection ${\cal{D}}$, and consequently its holonomy is
a subgroup of $Sim(3)$ \cite{Gibbons:2007}.

\section*{Acknowledgments}
The work of WS was supported in part by the National Science Foundation under grant number PHY-0903134. JG is supported by the EPSRC grant EP/F069774/1. AP is supported by a C.S.I.C. scholarship JAEPre-07-00176. AP would like to thank Patrick Meessen and Tom\'as Ort\'in for very useful discussions and technical guiding, as well as the \emph{Department of Fundamental Physics} at the Chalmers Tegniska H\"ogskola, and in particular Ulf Gran, for a research stay during the spring/ summer of 2010.

\renewcommand{\theequation}{A-\arabic{equation}}
% redefine the command that creates the equation no.
\setcounter{equation}{0} % reset counter

\end{document}